\documentclass[
  ,draft            
  ]
  {aipproc}

\layoutstyle{6x9}

\begin{document}

\title{On the Stability of a Planetary System Embedded in the $\beta$
Pictoris Debris Disk}

\author{Jared H. Crossley}{
  address={New Mexico Institute of Mining and Technology,
         801 Leroy Place, Socorro, NM 87801}
}

\author{Nader Haghighipour}{
  address={Department of Terrestrial Magnetism and 
NASA Astrobiology Institute, \\Carnegie Institution of
Washington, 5241 Broad Branch Road, Washington, DC 20015}
}

\begin{abstract}
It has recently been stated that the
warp structure observed around the star $\beta$ Pictoris
may be due to four planets embedded in its debris disk
\cite{wj03}. It, therefore, becomes important
to investigate for what range of parameters,
and for how long such a multibody system will be dynamically
stable. We present the results of the numerical integration
of the suggested planetary system for different values
of the mass and radii of the planets, and their relative
positions and velocities. We also present
a statistical analysis of the results in an effort to
understand the relation between the different regions of
the parameter-space of the system and the duration of the
orbital stability of the embedded planets.
\end{abstract}

\maketitle


\section{INTRODUCTION}

Beta Pictoris, a type A5 IV star 
at a distance of $\sim$19 pc from the Earth, is
one of the youngest close stars
with an approximate age of  $12^{+8}_{-4}$ Myr
\cite{zuckerman01}. Observational evidence indicates
that this star is surrounded by
a planetary debris disk \cite{st84,ho85, weinberger03}.
The close proximity of such a young
circumstellar disk has made $\beta$ Pictoris an ideal candidate 
for the study of the evolution of protoplanetary disks
and planetary system formation.

In the past several years, there have been a number of reports of
the detection of symmetric warps in the $\beta$ Pictoris debris disk
\cite{wj03, weinberger03,burrows95, mouillet97, heap00}. Models
have best explained these warps as gravitational perturbations 
caused by planetary
companions \cite{burrows95, mouillet97, heap00}. Among these reports,
the disk warps discovered by \citet{wj03} were explained as the edge-on
projection
of debris rings orbiting the central star.  Models of the flux density
allow some
parameters of these rings to be determined via $\chi^2$ fitting.
\citet{wj03} proposed
a multiple planet system to account for ring formation, 
noting that all adjacent rings
are in mean-motion resonances.

In consideration of these findings, we undertake here a study of the
dynamical evolution of
a multibody system similar to that proposed by \citet{wj03}.  We perform a
statistical analysis
of the stability of randomly generated systems within a portion of the
total available
parameter-space and briefly analyze the relationship
between the parameters of the system and its orbital stability.

\section{NUMERICAL ANALYSIS}
\label{system}

The planetary system proposed by \citet{wj03}
consists of four planets with
radial distances and orbital inclinations equal to those of their
corresponding warps \cite[see][Table 1]{wj03}.
To study the dynamics of this planetary system, we explore a
parameter-space which includes the mass number of the four planets,
their radii\footnote{
Planets' radii were needed to calculate their Hill's radii.},
and their positions and velocities.
The mass and radius of $\beta$ Pictoris
are taken to be $2.0\,M_\odot$ and $1.9\,R_\odot$, 
respectively \cite{co96}.

We consider planets with masses randomly chosen between
one to three Jupiter-masses.
For each value of the mass of a planet, we calculate its radius 
assuming an average density  equal to that of Jupiter
(1.33 gcm$^{-3}$).  We also assume that 
all planets are initially on direct Keplerian circular orbits
and their orbital phases are chosen
randomly from the range $0^\circ \le \phi \le 360^\circ$.

To explore the orbital stability of this planetary system, 
we integrated the system 
for 50 Myr using Mercury Integrator Package VI \cite{chmi99}. 
We considered the system to be stable if no planet came closer than
three Hill's radii or obtained a radial distance larger than 1000 AU.

We ran a total of 20457 simulations using randomly generated values
of planet masses and initial orbital phases.  Of this total, 14409
simulations used unique parameter sets. The remaining 6048 systems were
exact replications of the systems in the unique set---a consequence of
the random number generation routine.  The duplicate systems have been
kept for the sake of statistical analysis, since they are randomly 
distributed throughout the phase space.

Table 1 shows the statistical data for all simulations grouped in five
10 Myr intervals. The middle two columns show the number and percentage of
simulations that became unstable within their corresponding time intervals.
The majority of the randomly generated systems became 
unstable at early stages of the integration, with their number decreasing 
as time increases. The rightmost column shows the percentage of the systems 
remaining stable beyond their respective time interval.  
As shown here, approximately 41\% of the systems remained stable
after the first 10 Myrs. From these systems, 6.7\% were still stable
after 20 Myr from the beginning of the integrations--the 
upper estimate of the lifetime of $\beta$ Pictoris (20 Myr).
During the last 10 Myr, only 0.1\% of all systems were still stbale.

\begin{table}
\caption{Statistical data on five-body system stability.}
\begin{tabular}{rrrr}
  \hline
  \tablehead{1}{c}{b}{Time (Myr)}
  &\tablehead{1}{c}{b}{Number of \\Unstable Systems}
  &\tablehead{1}{c}{b}{Percentage of \\Unstable Systems}
  &\tablehead{1}{c}{b}{Percentage of Remaining \\Stable Systems} \\
  \hline
  0--10      & 12100  & 59.1     & 40.9   \\
  10--20     & 6980   & 34.1     & 6.7    \\
  20--30     & 1152   & 5.6      & 1.1    \\
  30--40     & 194      & 0.9      & 0.2    \\
  40--50     & 16       & 0.1      & 0.1 \\
  \hline
\end{tabular}
\end{table}

\begin{figure}[b]
  \begin{minipage}[t]{2cm}
  \rotatebox{-90}{ \scalebox{0.4}
                   {\includegraphics{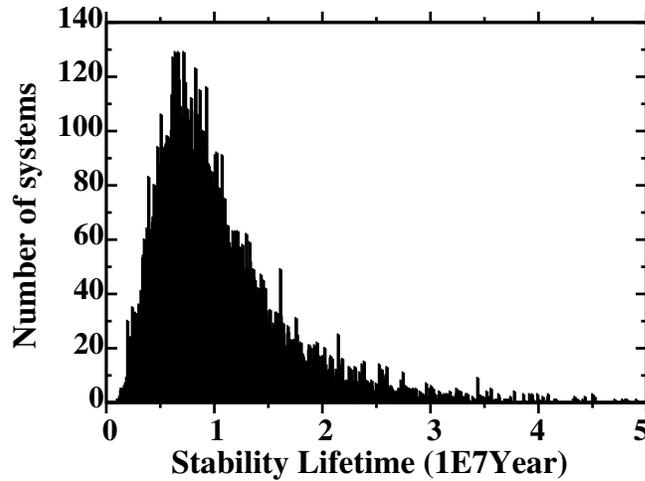}}}

  \caption{Histogram showing the number of systems that became unstable in
      $10^5$\,yr intervals.  A randomly chosen system is most likely to
      become unstable near 7\,Myr.  No system in this sample became unstable in
      less than 1\,Myr. The 15 systems that remained stable for the 
      entire 50 Myr integration are not shown here.
  \label{stable_hist} }
  \end{minipage}
  \hfill
\end{figure}

\begin{figure}[htb]
  \begin{minipage}[t]{2.5cm}
  \rotatebox{-90}{ \scalebox{0.4}
                  { \includegraphics{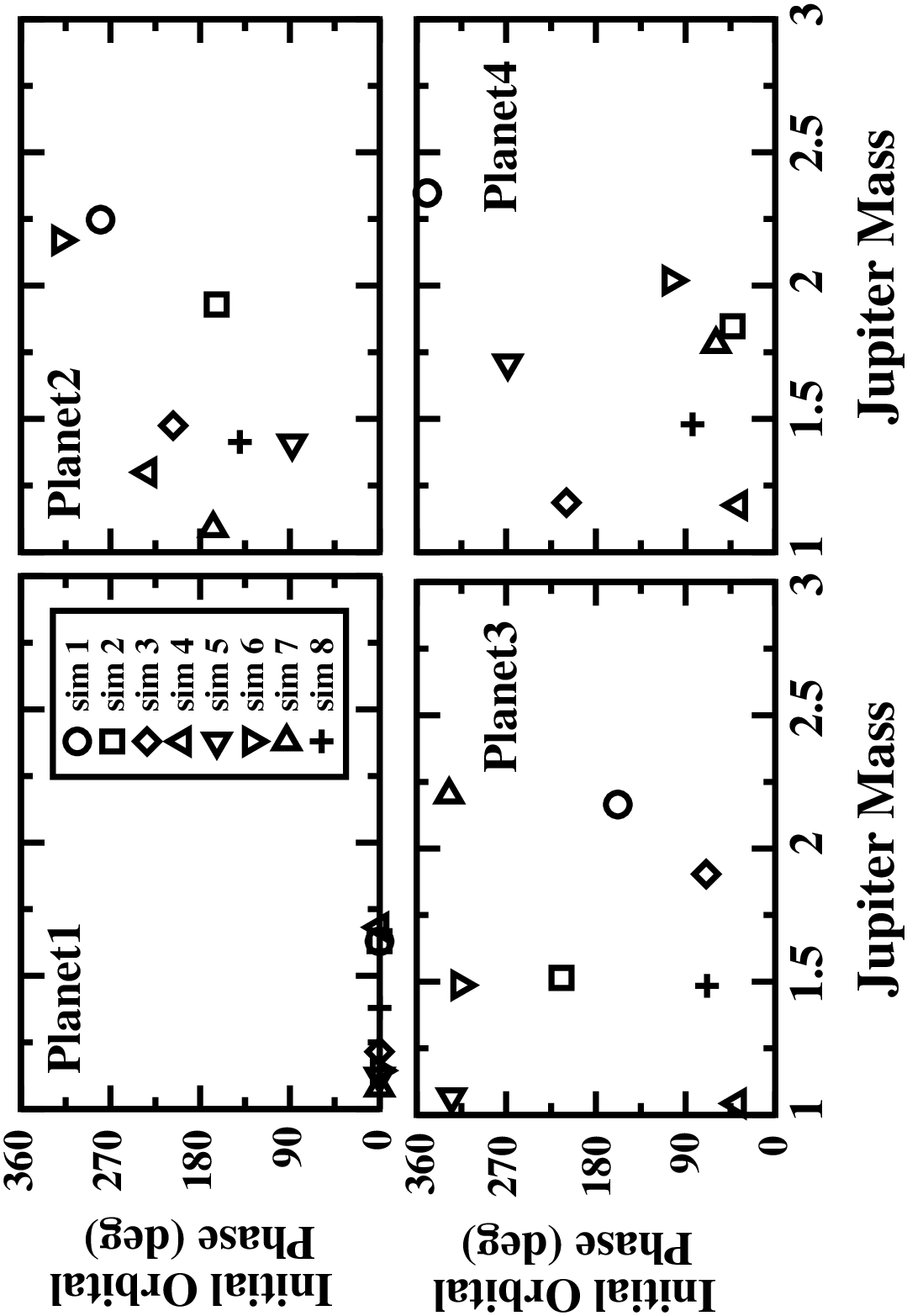}}}
  \caption{Mass versus initial phase for planets in the 8 unique systems
that remained stable for the entire 50\,Myr integration. Note that none
of these planets have masses greater than 2.4 Jupite-mass.
  \label{mass_vs_phase} }
  \end{minipage}
\end{figure}

Figure \ref{stable_hist} shows how stability lifetimes are distributed
across a random
sampling of the parameter-space.  This histogram shows the number of systems
that became unstable in $10^5$ year intervals.  It can be seen from this
figure that for a system chosen randomly within our assumed
parameter-space, it is most probable that the stability lifetime 
of the system is approximately 7\,Myr.  It is also seen that no system
became unstable at an age of less than 1.0\,Myr. We note
that 15 systems remained stable for the entire 50\,Myr
and are therefore not accounted for in the histogram.

In a preliminary attempt to determine the relationship between 
initial conditions
and stability lifetime, we have plotted
the mass of each planet versus its initial phase-angle for those systems
that remained stable for 50 Myr (Fig. 2).  
The initial phase of
planet 1 was set to $0^\circ$ for all simulations.
 It is interesting to note that none of these systems
contain a planet with a mass greater than $2.4\,M_J$.

\section{CONCLUSIONS}
\label{conclusions}

We have analyzed the stability of the proposed five-body planetary system
embedded in the $\beta$ Pictoris debris disk \citep{wj03} for
over 14000 initial conditions. Our results indicate that 
the majority of systems became unstable between a time of 1 to 10
million years.
There were only 8 unique simulations that remained stable for the entire
50\,Myr integration time. These systems contained planets with masses
less than 2.4 times the mass of Jupiter.


\begin{theacknowledgments}
This work is partially supported by the Carnegie Institution
of Washington Internship Program, and also an REU Site for 
Undergraduate Research Training in Geoscience, NSFEAR-0097569
for JHC, and NASA Origins of the Solar System Program under
grant NAG5-11569, and also the NASA Astrobiology Institute
under Cooperative Agreement NCC2-1056 for NH.
\end{theacknowledgments}

\bibliographystyle{aipproc}

\begin{thebibliography}
\expandafter\ifx\csname natexlab\endcsname\relax\def\natexlab#1{#1}\fi
\providecommand{\enquote}[1]{``#1''}
\expandafter\ifx\csname url\endcsname\relax
  \def\url#1{\texttt{#1}}\fi
\expandafter\ifx\csname urlprefix\endcsname\relax\def\urlprefix{URL }\fi


\bibitem[Wahhaj et al.(2003)]{wj03}
Wahhaj, Z., Koerner, D. W., Ressler, M. E.,
Werner, M. W., Backman, D. E., and Sargent, A. I.,  
\emph{Astroph. J.}, \textbf{584}, L27--L31 (2003).
\bibitem[Zuckerman et al.(2001)]{zuckerman01}
Zuckerman, B., Song, I., Bessel, M. S., and
Webb, R. A., \emph{Astroph. J.}, \textbf{562}, L87--L90 (2001).
\bibitem[Smith and Terrile(1984)]{st84}
Smith, B. A., and Terrile, R. J., \emph{Science}, 
\textbf{226}, 1421--1424 (1984).
\bibitem[Hobbs et al. (1985)]{ho85}
Hobbs, L. M., Vidal-Madjar, A., Ferlet, R., Albsert, C. E.,
and Gry, C., \emph{Astroph. J.}, \textbf{293}, L29--L33 (1985). 
\bibitem[Weinberger et al.(2003)]{weinberger03}
Weinberger, A. J., Becklin, E. E.,
and Zuckerman, B., \emph{Astroph. J.}, \textbf{584}, L33--L37 (2003).
\bibitem[Burrows, Krist, and Stapelfeldt(1995)]{burrows95} 
Burrows, C. J., Krist, J. E., and Stapelfeldt, K. R., 
\emph{Bul. Am. Astro. Soc.}, \textbf{27}, 1329 (1995).
\bibitem[Mouillet et al.(1997)]{mouillet97} 
Mouillet, D., Larwood, J. D., Papaloizou, J. C. B., and
Lagrange A. M., \emph{Month. Not. Roy. Astron. Soc.}, 
\textbf{292}, 896--904 (1997).
\bibitem[Heap et al.(2000)]{heap00}
Heap, S. R., Lindler, D. J., Lanz, T. M., 
Cornett, R. H., Hubeny, I., Maran, S. P., and Woodgate, B.,
\emph{Astroph. J.}, \textbf{539}, 435--444 (2000).
\bibitem[Carroll and Ostlie(1996)]{co96}
Carroll, B. W., and Ostlie, D. A., 
\emph{An Introduction to Modern Astrophysics}, 
Addison-Wesley, New York, PP. A--13 (1996).
\bibitem[Chambers (1999)]{chmi99}
Chambers, J. E., \emph{Month. Not. Roy. Astron. Soc.},
\textbf{304}, 793--799 (1999).




\end{thebibliography}

\end{document}